 \documentclass[preprint,review,12pt]{elsarticle}

\usepackage{lineno}
\usepackage{amsmath}
\usepackage{amssymb}
\usepackage{amsthm}
\usepackage{textcomp}
\usepackage{color}
\usepackage{graphics}
\usepackage{graphicx}
\usepackage[latin1]{inputenc}
\usepackage[T1]{fontenc}
\usepackage{multirow}
\usepackage{threeparttable}

\biboptions{sort&compress}

\begin{document}

\begin{frontmatter}



\title{Low-dimensional modelling of flame dynamics in heated microchannels}

\author[ijlra,roma]{Federico ~Bianco} 
\ead{fedbianco@gmail.com}
\author[ijlra]{Sergio ~Chibbaro \corref{cor1}}
\ead{chibbaro@ida.upmc.fr}
\author[ijlra]{Guillaume ~Legros}
\ead{guillaume.legros@upmc.fr}
\address[ijlra]{Sorbonne Universit\'es, UPMC Univ Paris 06, CNRS, UMR7190, Institut Jean Le Rond d'Alembert, 
F-75005 Paris, France.}
\address[roma]{Dipartimento di Fisica, Universit\'a ``La Sapienza'', \\
Piazzale Aldo Moro 2, I-00185 Roma, Italy}

\cortext[cor1]{Corresponding author}

\begin{abstract}
This paper presents simulations of stoichiometric methane/air premixed flames into a microchannel at atmospheric pressure. These simulations result from numerical resolutions of low-order models. Indeed, combustion control into microchannels would be allowed by fast simulations that in turn enable real-time adjustments of the device's parameters.
Former experimental studies reported the occurrence of a Flame Repetitive Extinction/Ignition (FREI) phenomenon provided that a temperature gradient is sustained at the channel's walls. Conducting unsteady one-dimensional simulations including complex chemistry, a late numerical study tried to explain the occurrence of this phenomenon. The present study therefore explores low-order models that potentially reproduce the FREI phenomenon. Provided a calibration of some empirical constants, an unsteady two-dimensional model including one-step chemical reaction is shown to decently reproduce the FREI regime all along the range of mixture flow rates investigated by the experimental studies. Complementing the aforementioned numerical study,  furthermore, when the channel's diameter is varied, the two-dimensional model unveils an unstable regime that a one-dimensional model cannot capture. As two-dimensional hydrodynamics appears to play a key role into the flame's dynamics, therefore the heat rate released by the microcombustor, one-dimensional models are not believed to deliver an adequate strategy of combustion control into such microchannels.
      
\end{abstract}


\end{frontmatter}


\section{Introduction}

Significant efforts have been lately devoted to the design of microcombustor technologies, 
that may enable the development of micro power generation devices with low weight and long
 life \cite{maruta11}. The potential of such devices is supported by the energy density of
 hydrocarbon fuels which is almost two orders of magnitude higher than that of modern batteries \cite{fernandezpello02}.
Nonetheless, the practical performance of a microcombustor is especially constrained by
 both the low overall efficiency and the narrow range of operational conditions.

As the ratio of the reacting volume to the wall surface decreases, the proximity of the reacting volume from the colder walls can lead to higher heat losses, therefore extended flame thermal quenching.
In addition, heterogeneous chemical reactions may
 occur at the walls, possibly contributing to radical quenching of the flame \cite{pizza09}.

Heat-recirculating combustors, such as the double-spiral counter-current Swiss roll \cite{lloyd74,chen11}, have been developed to overcome this trouble. In such a
 device, thermal energy flows from combustion products to reactants through
 a wall, i.e. without any mass transfer that would induce dilution of reactants. The incoming
 cold reactants enthalpy is then increased and may sustain combustion under conditions that would 
lead to extinction without recirculation.

However, the premixed flame propagation in such ducts may exhibit different combustion modes, 
which can make any microcombustor's design streneous. Sustaining steady conditions, some studies 
reported steady mild -or flameless- combustion \cite{maruta05,pizza08a}, non-axisymmetric flames
 in circular ducts \cite{kurdyumov02,kurdyumov07,tsai08}, asymmetric flames in planar channels \cite{pizza08a,dogwiler98}, 
and tulip flames \cite{dunn86,bychkov07}. Unsteady behaviors also revealed, 
such as periodic flame repetitive ignition/extinction (FREI) \cite{maruta05,richecoeur05,jackson07},
 oscillating flames \cite{maruta05,pizza08a,kurdyumov08} and spinning flames \cite{kwon96,xu07}. 
Therefore, the industrial development of reliable microcombustors requires fundamental 
understandings of premixed flame propagation at small scales.

Within the context of fundamental works on combustion at small scales, 
the studies of reactive flows in a straight heated channel with an inner diameter smaller than
 the ordinary quenching diameter at ambient conditions were shown to provide
 meaningful contributions \cite{maruta05,richecoeur05,fan09,pizza10}.
 Provided that a temperature gradient at the channel's wall can be controlled,
 such a configuration lately gave significant insights into both  the 
ignition and combustion characteristics of alternative fuels \cite{yamamoto11} 
and the physico-chemical processes that govern the repetitive 
ignition/extinction regime \cite{nakamura12}. Thus, strategies combining flow, 
thermal and chemical managements are required to establish stable 
combustion in micro and mesoscale devices.
As an illustration, Pizza et al \cite{pizza09} showed that some
undesirable unsteady combustion modes may be suppressed when
 applying a predetermined catalyst loading on the channel walls. 
The subsequent increased catalytic reactivity may indeed 
result in a decreased sensitivity of the homogeneous ignition
 distance to small perturbations of the gaseous reactivity.
Nevertheless, some physical mechanisms underlying instabilities remain to be understood
for the relevant range of parameters.

While low-order modelling of reactive flows in micro-channels enables relevant simulations of some regimes constituting flame dynamics, limitations of present models, especially one-dimensional, are to be assessed.
For instance, while Nakamura et al \cite{nakamura12} qualitatively investigated the phenomena driving the FREI regime, further questioning the scope of such low-order models is required when developing predictive tools aiming at the control of micro-combustors. 

To this end, an unsteady two-dimensional model incorporating a one-step chemical kinetics
 is here introduced. The flow is imposed and the coupled energy and species equations are numerically solved. A calibration procedure then allows the subsequent computations to fairly capture the FREI regime, suggesting that a thermal management of the microcombustor's wall could provide with a relevant strategy of combustion control. Further simulations explore other regimes that reveal when varying the fresh gases velocity at constant channel diameter. Interestingly, when exploring the parameters' space in terms of fresh gases velocity and channel diameter, the two-dimensional model eventually unveils a region of instability while a one-dimensional simulation predict a stable regime.

\begin{figure}[ht!]
\centering
\includegraphics[scale=0.5]{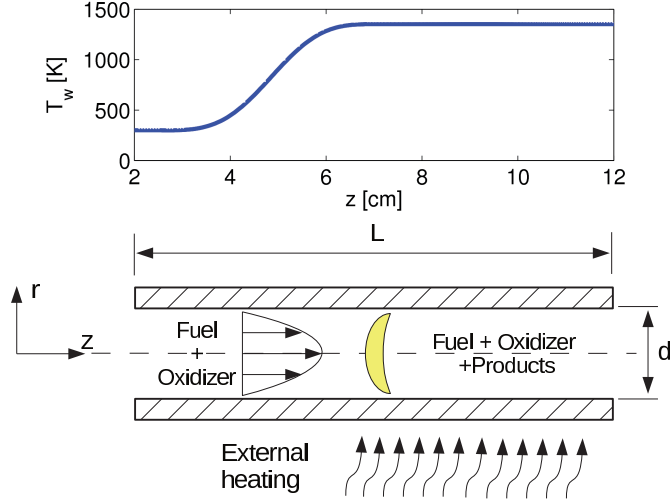}
\caption{Schematic of the geometrical configuration investigated. The evolution along the channel's axis of the temperature T$_\mathrm{w}$(z) imposed at the wall is shown at the top.}
\label{fig:sketch}
\end{figure}

\section{Theoretical models and numerical resolution}

\subsection{Configuration and limitations}

		We consider a microchannel whose diameter is $d$, as shown on the schematic in Fig.				\ref{fig:sketch}. The coordinate along the channel's axis is $z$ and the distance from this
		axis is $r$.

		Within the range of parameters investigated, the flow is laminar and steady.
		At the upstream boundary of the domain, the flow is composed of methane and air in 				stoichiometric proportions.
		The flow then experiences an increase of the  wall temperature $T_w$ along the 					streamwise coordinate $z$, therefore can possibly ignite.
		This configuration matches as much as possible
		the experimental setup described by Tsuboi et al \cite{Tsuboi09}.

		To set a low-order model for this configuration, the following assumptions are stated:
		\begin{enumerate}[({hyp.}1)] 
			\item \label{hyp1} the Mach number of the flow is low;
			\item \label{hyp2} the radial component $u_r$ of the flow velocity is zero while the 					axial 	velocity $u_z(r,z)$ is two-dimensional and axially symmetric;
			\item \label{hyp3} Lewis number is unity;
			\item \label{hyp4} Fick's law governs the diffusion velocities;
			\item \label{hyp5} the thermophysical properties, i.e. mass diffusion coefficient $D$, viscosity $\eta$, and the specific heat $C_p$, are 				species independent constants;
			\item \label{hyp6} this flow is composed of perfect gases and has a constant density 					$\rho$;
			\item \label{hyp7} the chemical kinetics is restricted to a single reaction including 					CH$_4$, O$_2$, CO$_2$, and H$_2$O.
		\end{enumerate}
		
		The range of mean axial velocity $U_0$ investigated, i.e. $0 \le U_0 \le 100$ cm/s,
		clearly allows assumption (hyp.\ref{hyp1}) to be stated.

		The Reynolds number is then low enough to guarantee Hagen-Poiseuille law.
		Indeed, the simulations reported hereafter do not intend to reproduce high
		Reynolds number flows. Therefore, investigating the dynamic regimes leading to 					asymmetric flames, such as those computed by Pizza et al \cite{pizza08a}, is beyond the 				scope of the present study. Furthermore, at hydrodynamic conditions similar to those 				explored here, Pizza et al. \cite{pizza10} reported velocity profiles that are in agreement with 			assumption (hyp.\ref{hyp2}) (see Fig.5 in \cite{pizza10}). Thus, this assumption is not 			considered restrictive.

		For stoichiometric mixtures of methane and air, Lewis number is closed to unity. 					Therefore, assumption (hyp.\ref{hyp3}) does not represent a significant limitation. 

		Fick's law is exact for binary mixtures, which makes assumption (hyp.\ref{hyp4}) especially 			relevant for low-order modelling where chemical kinetics only include few species 					\cite{poinsot}.

		Thus, the low-order characteristics of the following model can essentially be attributed to 				assumptions (hyp.\ref{hyp5}) to (hyp.\ref{hyp7}). Nonetheless, with these assumptions, some 				theoretical studies, such as that of Bai et al. \cite{bai13}, captured the structure of premixed 			flames in microchannels with some success.

\subsection{Two-dimensional model}		
		
Given these assumptions, momentum equation is decoupled from energy and 						species equations. The problem is then fully described by the following set of partial 					differential equations~\cite{Peters}:
\begin{align}
\frac{\partial{Y_k}}{\partial t} + u_r \frac{\partial{Y_k}}{\partial r}+u_z \frac{\partial{Y_k}}{\partial z} = D
 \left[ \frac{1}{r}\frac {\partial}{\partial r} \left( r\frac {\partial{Y_k}}{\partial r} \right)
 		 +    \frac{\partial^2{Y_k}}{\partial z^2}  \right]  + \frac{\dot\omega_k}{\rho} \nonumber \\
\frac{\partial{T}}{\partial t} + u_r \frac{\partial{T}}{\partial r}+u_z \frac{\partial{T}}{\partial z} = D
 		\left[ \frac{1}{r}\frac {\partial}{\partial r} \left( r\frac {\partial{T}}{\partial r} \right)
  		+    \frac{\partial^2{T}}{\partial z^2}  \right]  + \frac{\dot\omega_T}{\rho C_p}
		\label{sistema2D}
\end{align}
where  $t$ is time, $Y_k$ the mass fraction of the $k^{th}$ species considered, $\dot\omega_k$ the production rate of this species, $T$ the temperature of the mixture, and $\dot\omega_T$ the heat release rate.

		Since momentum equation is decoupled from Eqs.\eqref{sistema2D}, the velocity field 				inside the tube evolves independently.
		Given the pressure gradient $\frac{\partial p}{\partial z}$ in the direction of the flow, the  				Hagen-Poiseuille law can be expressed as follows:
		\begin{align}
		u_r&=0\nonumber\\
		u_z&=-\frac{1}{4\eta}\frac{\partial p}{\partial z}(R^{2}-r^{2})=-\frac{2 U_0}{R^2}(R^{2}-r^{2}),
		\end{align}

		Furthermore, chemical kinetics is reduced to a one step irreversible reaction:  
		 \begin{equation}
		\label{eq:stoechio}
		CH_4+2(O_2+3.76N_2) \rightarrow CO_2 + 2H_2O + 7.52 N_2  \nonumber
 		\end{equation}

		The forward reaction rate of methane oxidation in air can be evaluated with the following 				global law:
		\begin{equation}
		 K_f=A T^n \left( \frac{\rho Y_{CH_4}}{W_{CH_4}}\right)^a  \left( \frac{\rho Y_{O_2}}					{W_{O_2}}\right)^b \exp{\left( -\frac{T_a}{T}\right)}
 		\label{rateWest}
		\end{equation}
		\noindent where $A$ is the pre-exponential factor, and $W_{CH_4}$ and $W_{O_2}$ are the 			molecular weight of methane and oxygen, respectively. According to Westbrook and Dryer 				\cite{westbrook1981simplified}  $n=0$, 			$a=-0.3$, $b=1.3$, and $T_a=24200 K$. 

		Since the change in composition of the mixture is then determined by a single
		step irreversible reaction and the density of the mixture 
		is supposed to be constant while species diffuse at the same 
		velocity, the composition of the mixture can be expressed everywhere
		as function of only one species. 

		The set of equations (\ref{sistema2D}) is then reduced to a system of 				two equations: 
		\begin{align}
		\frac{\partial{Y_{CH_4}}}{\partial t} + u_z \frac{\partial{Y_{CH_4}}}{\partial z} &= D
		 \left[ \frac{1}{r}\frac {\partial}{\partial r} \left( r\frac {\partial{Y_{CH_4}}}{\partial r} \right)
  		+    \frac{\partial^2{Y_{CH_4}}}{\partial z^2}  \right]  + S_{Y} \nonumber \\ 						
		\frac{\partial{T}}{\partial t} + u_z \frac{\partial{T}}{\partial z} &= D
 		\left[ \frac{1}{r}\frac {\partial}{\partial r} \left( r\frac {\partial{T}}{\partial r} \right)
  		+    \frac{\partial^2{T}}{\partial z^2}  \right]  + S_{T}
		\label{sistema2Drisolto}
		\end{align} 

		$s$  being the mass stoichiometric oxygen/methane ratio,
		the oxygen mass fraction $Y_{O_2}$ can be replaced in Eq.\eqref{rateWest} by $sY_{CH_4}$. If we then collect 
		all constant parameters in $A^*$, the rate of methane consumption and the specific heat released
 		by reaction are respectively as follows:
		\begin{align}
		S_Y&=\frac{\dot\omega_{CH_4}}{\rho}=\frac{\nu_{CH_4}W_{CH_4}}{\rho}K_f=-A^* Y_{CH_4} 			\exp{\left( -\frac{T_a}{T}\right)} \nonumber \\
		S_T&=\frac{\dot\omega_{T}}{\rho C_p}=-\frac{Q}{C_p} S_Y
		\end{align}
		\noindent where $Q$ is the heat of combustion and $\nu_{CH_4}$ the molecular stoichiometric coefficient  in Eq.(\ref{eq:stoechio}), i.e. -1.

		For methane/air stoichiometric mixture $s$=4,  
		while $Q/C_p=35000 K$ 
		is required to match the flame adiabatic temperature at constant pressure ($p_0$=1 bar , 				$T_0$=300 K).
		Only two parameters, i.e. $D$ and $A^*$, remain undetermined. These will be calibrated to 				quantitatively match the dynamic response of the system with the experimental behavior.

		The computational domain is defined as $0\le z\le L$ and $0 \le r \le R$, with $R$=$d/2$.
		The temperature at the wall $T(R,z)$ is imposed and its profile is taken from Ref. 					\citealp{Tsuboi09} (see Fig.\ref{fig:sketch}).
Inflow conditions for temperature and methane mass fraction at $z=0$ are as follows:
\begin{equation}
T(r,0)=T^0=300 K~,~Y_{CH_4}(r,0)=Y_{CH_4}^0=0.055 
\end{equation}
Outflow conditions at $z=L$ are zero gradient ones:
\begin{equation}
\frac{\partial{T(r,L)}}{\partial z}=0~,~\frac{\partial{Y_{CH_4}(r,L)}}{\partial z}=0
\end{equation}
On the channel's axis ($r=0$), conditions of symmetry are forced:
\begin{equation}
\frac{\partial{T(0,z)}}{\partial r}=0~,~\frac{\partial{Y_{CH_4}(0,z)}}{\partial r}=0
\end{equation}

\subsection{One-dimensional model}	

Since the diameter $d$ of the channel is small, compared to its characteristic length $L$,
 the problem could be further approximated by a one-dimensional model. 
Some authors \cite{maruta05,Tsuboi09,nakamura12} have tried to reproduce microcombustors dynamics considering
 a plug-flow approximation. For a matter of quantitative assessment of the performances provided by both kinds of model, any quantity inferred from the two-dimensional model can be averaged  over the channel's section $S$ as follows:
\begin{equation}
\bar f(z,t)=\frac{2\pi}{S}\int_{0}^R f(r,z,t) r dr 
\end{equation}

Following these works\cite{maruta05,Tsuboi09,nakamura12}, a plug-flow 1-D model, 
governing the spatial and temporal evolution of mean temperature and methane mass fraction, can be defined as follows:
\begin{align}
\frac{\partial{\bar Y_{CH_4}}}{\partial t} 
+ U_0 \frac{\partial{\bar Y_{CH_4}}}{\partial z} 
&= 
D \frac{\partial^2{\bar Y_{CH_4}}}{\partial z^2}
+ S_{\bar{Y}}\nonumber \\
\frac{\partial{\bar T}}{\partial t} 
+ U_0 \frac{\partial{\bar T}}{\partial z} 
&= 
D \frac{\partial^2{\bar T}}{\partial z^2} +S_{\bar{T}}  -H_l 
\label{sistema1D}
\end{align}
where the rate of methane consumption and specific heat released, are
function of mean temperature and mass fraction only
\begin{align}
S_{\bar{Y}}&=- A^* \bar{Y}_{CH_4} \exp{\left( -\frac{T_a}{\bar T}\right)} \nonumber \\
S_{\bar{T}}&=-\frac{Q}{C_p}\bar{S}_Y,
\end{align}

$U_0$ is the mean flow velocity. To allow direct comparison, the values of $D$, $A^*$, and $Q/Cp$ are identical to those of the 2D model.

The heat rate $H_l$ exchanged between the reacting mixture 
and the walls is modeled by an ad-hoc term which,
 assuming unit Lewis number, is of the form~\cite{maruta05,nakamura12,Tsuboi09}:
\begin{equation}
H_l=\frac{4DNu}{d^2} (\bar T-T_w)
\label{eq:nusselt}
\end{equation}
where $Nu$ is
the Nusselt number which is the only free parameter left.

\subsection{Numerical tools}
The numerical resolution of models  (\ref{sistema2Drisolto}) and (\ref{sistema1D}) is not straightforward.
 The system of equations is stiff due to the presence of the
 reaction terms. Indeed, the chemical characteristic time is sufficiently small as compared to
 the transport characteristic time. Therefore, the numerical time 
step needs to be sufficently small to describe accurately the fast heat released by reaction while,
 especially in the case of unstable dynamics, the solution needs to be observed for a 
sufficiently long period of time. Moreover, when the fresh mixture coming from  the inlet is ignited,
 there exists a flame, in a certain position of the domain, at time $t$, where most of the chemical
 reaction takes place.  Within this area, rapid variations of temperature and mass fraction
 occur and the mesh should be sufficiently refined to capture these strong gradients.  
If we consider the configuration exhibited in Fig.\ref{fig:sketch}, the length of the cylindrical domain is $L$=10 cm while 
the diameter is of the order of $2$ mm. The typical flame thickness of 
a premixed methane/air flame is of the order of 0.2 mm~\cite{flamethickness}. 
To reduce the total number of grid points, a moving mesh method has been implemented to dynamically refine the mesh around the flame.

\subsubsection{Space and time discretization}

\begin{figure}[!ht]
 \centering{
 \begin{tabular}{c}
 \includegraphics[scale=0.4]{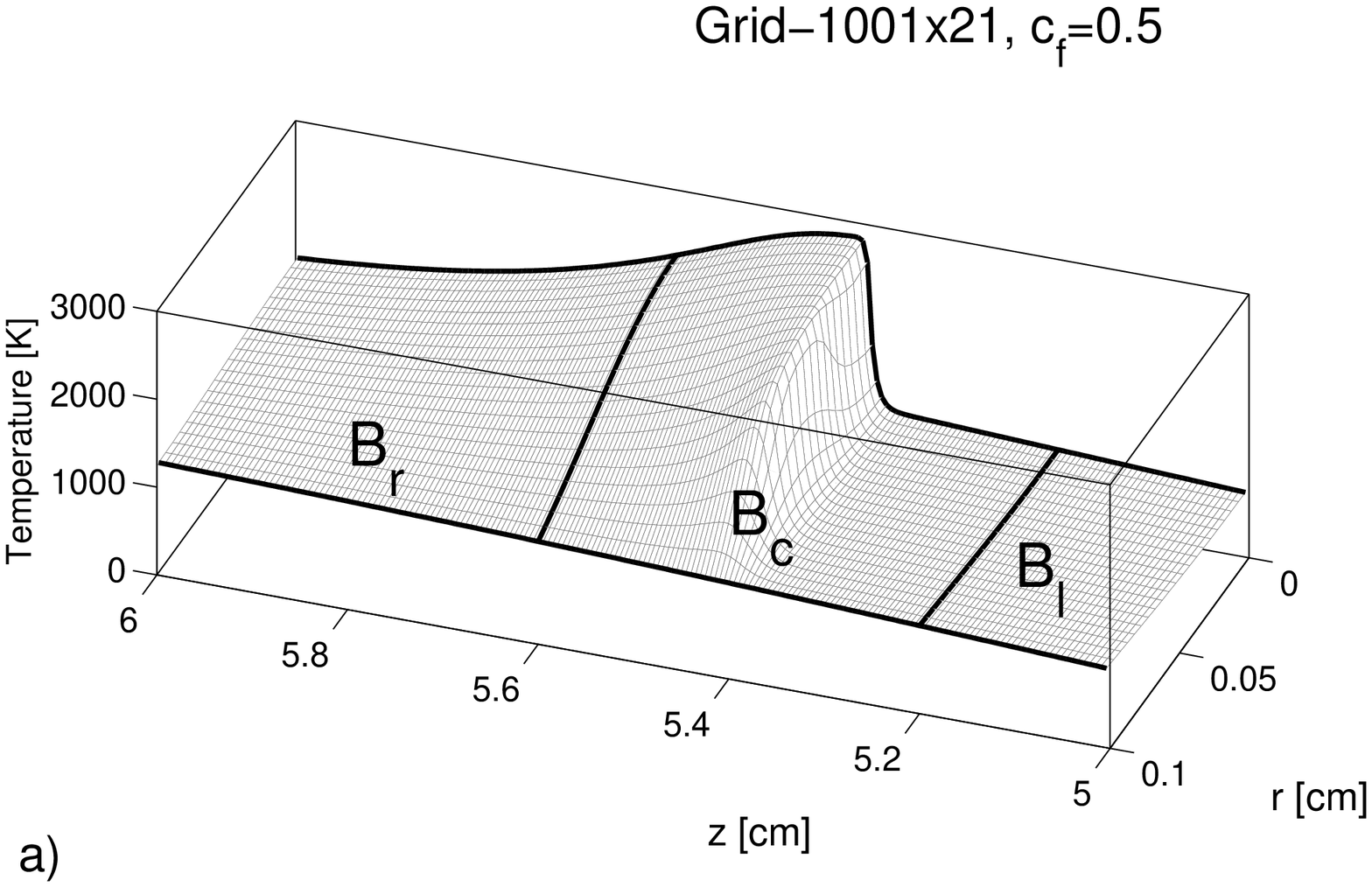} \\
\includegraphics[scale=0.4]{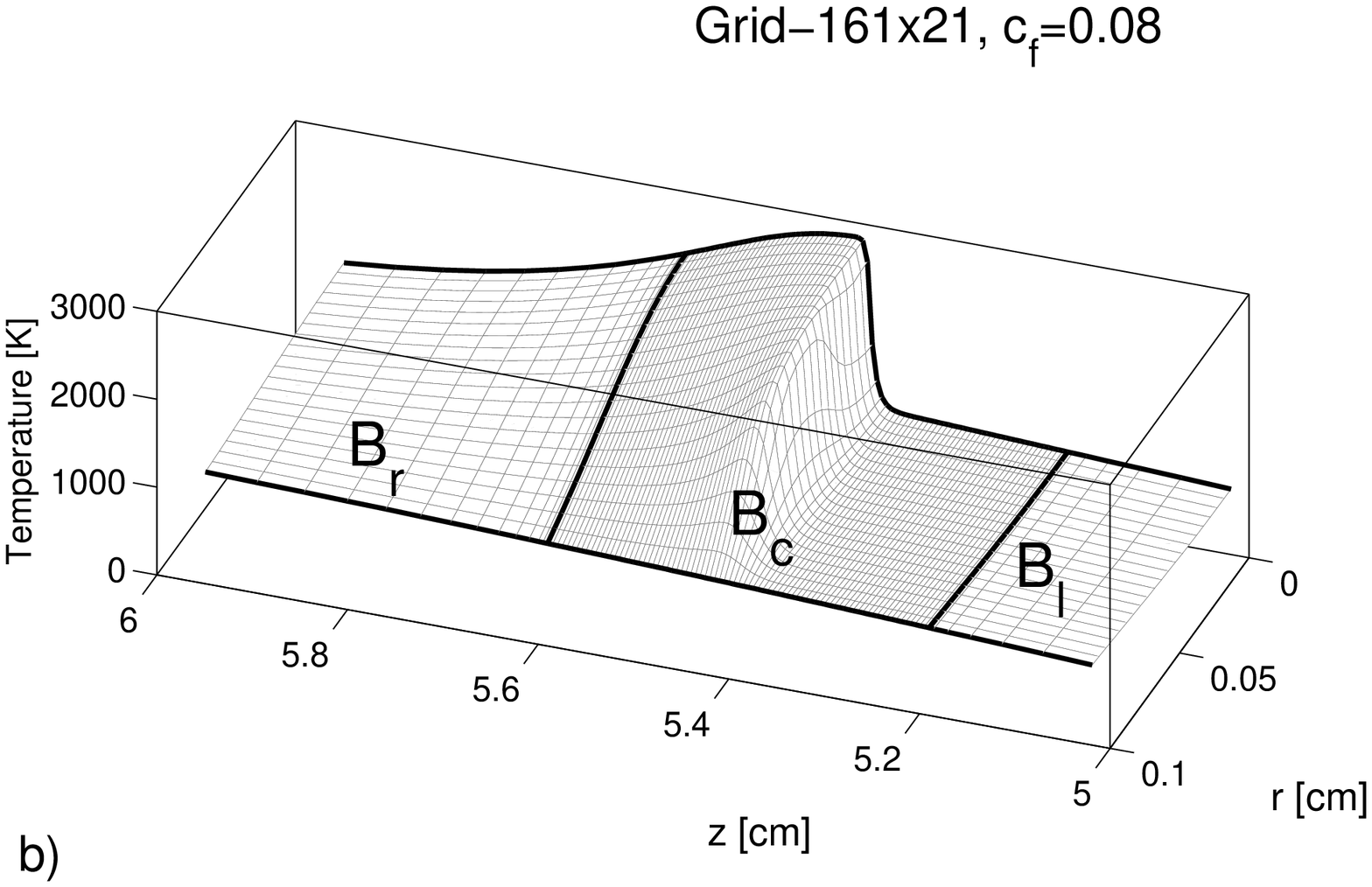} 
 \end{tabular}
 }
 \caption{ Examples of adaptive moving mesh. In panels (a,b) a snapshot of the temperature profile around the flame is shown. In the central moving subdomain ($B_{c}$), which contains the flame, the grid spacing along $z$ is constant and equal to $dz_{c}= c_{f}L/(N_{z}-1)$. In the left and right subdomains ($B_{l}$, $B_{r}$) the grid spacing varies from $dz_{c}$ to a maximum value at the tube ends. The grid spacing along $r$ is constant and equal to $dr=dz_{c}$. In panel (a) with a compressing factor $c_{f}=0.5$, $N_{z}=1001$ grid points along  $z$ are needed to guarantee a resolution of the grid $dz_{c}=0.005 [cm]$ around the flame.  In panel (b) with $c_{f}=0.08$, $N_{z}=161$  grid points are required to guarantee the same resolution. }
   \label{fig:mesh}
 \end{figure}

Let consider the 2D axially-symmetric  model  (\ref{sistema2Drisolto}) since the 1D case may be derived straightforward. The numerical domain is discretized with a cartesian grid of $N_{z} \times N_{r}$ grid points along the cylinder length and radius respectively, where:
\begin{eqnarray}
 z(i)\in[0,L]~~~i=1,\dots,N_{z} \nonumber \\ 
 r(j)\in[0,R]~~~j=1,\dots,N_{r}   
\end{eqnarray}
are the cylindrical coordinates of the generic grid point $(i,j)$. \\
 Many different strategies to locate the flame and consequently adapt the mesh may be considered. We found a good compromise, in terms of efficiency and coding, if we follow in time the value of the source term $S_{T}$ of Eq.~\eqref{sistema2Drisolto}.\\
 We know from simulations that a flame exists if somewhere in the domain the value of $S_{T}$ is greater than a threshold $S^{*}_{T}$ (here $S^{*}_{T}\approx1\cdot10^{5}$) and its position can be located by finding:
\begin{equation}
z^{*}(t)  \mid  S_{T}(z^{*},R/2,t)=\max\{ S_{T}(z,R/2,t) \} 
\end{equation}
If there exists a flame and mesh refinement  is required, the numerical domain is split in three blocks: $B^{l}$, $B^{c}$ and $B^{r}$ (see Fig.~\ref{fig:mesh}). The central moving block:
\begin{equation}
B^{c}:=[ 0 , R ]\times ] z^{*}(t)-z_{l} , z^{*}(t)+z_{r} [
\end{equation}
contains the flame and $N^{c}_{z}$ grid-points are dedicated to this portion of the domain such that the grid spacing inside the block is constant and equal to:
\begin{equation}
dz^{c}=c_{f} (L/(N_{z}-1))
\label{eq:cf}
\end{equation}
where $c_{f}<1$ is the coarse factor of the grid. The extent of the central block ($z_{l}+z_{r}$) has to be sufficiently large in order to contain the entire 2D flame and for most of the computations we have considered $z_{l}=0.2[cm]$, $z_{r}=0.2[cm]$.\\
The other  $N_{z}-N^{c}_{z}$ grid points are redistributed between the remaining blocks:
\begin{align}
B^{l}&:=[ 0 , R ]\times [0 , z^{*}(t)-z_{l} ] \nonumber \\
B^{r}&:=[ 0 , R ]\times [ z^{*}(t)+z_{r} , L ]
\end{align}
proportionally to their extent along $z$  ($L^{l}=z^{*}-z_{l}$ and $L^{r}=L-z_{r}-z^{*}$). The distance between two grid points ($dz^{l}$,$dz^{r}$) varies from a minimum value $dz_{c}$, adjacent to the central block, to a maximum value at the cylinder ends (to be determined) following a power law.\\
 The number of grid points along the radial direction $N_{r}$ is such that the grid spacing is constant and equal to:
\begin{equation}
dr=R/(N_{r}-1)
\end{equation}
Some examples of refined mesh are shown in Fig.~\ref{fig:mesh}.\\
If $S_{T}(r,z,t)\le S^{*}_{T}$, $\forall (r,t) \in B^{l}\cup B^{c}\cup B^{r}$, i.e. there is no flame in the domain, the cylinder is discretized
with a regular grid and the grid points are equally distributed so that:
\begin{align}
dr&=R/(N_{r}-1)\nonumber \\
dz&=L/(N_{z}-1)
\end{align}
Once a proper mesh is defined, the spatial derivatives in the inner domain ($2<i<N_{z}-1$,$2<j<N_{r}-2$) are computed via finite difference method in a central nine-points stencil. 
 Approaching the boundaries of the domain, if the computation of spatial derivatives is required, the stencil is changed from central to backward (or forward depending on the boundary approached) while the number of grid-points in the stencil is unchanged. \\ 
Every time the mesh is updated, the position $(z,r)$ of the generic grid point $(i,j)$, as well as the distance between grid points of the same stencil, changes. Therefore, finite difference weights have to be calculated every time a change in the grid occurs. An efficient algorithm for the calculation of the weights in finite difference formulas has been proposed by  Fornberg~\cite{Fornberg98,fornberg1988generation}. \\
For the time integration, we find convenient to use an explicit $4^{th}$ order Runge-Kutta method which is easy to implement and parallelize using OpenMP where, taking advantage of a multicore architecture, the evaluation of spatial derivatives, and the time marching, can be distributed over different threads which work simultaneously on different portions of the discrete domain $(i,j)$.\\
Finally, the Runge-Kutta  method has been validated by comparing the transient solution with an explicit Euler method with a significantly smaller time step while, the mesh refining algorithm has been validated by comparing the results with an equally spaced grid. Convergence of the solution is obtained if we consider a  grid spacing and a time step lower than $1.5\cdot10^{-3}[cm]$ (near flame) and $1\cdot10^{-6}[s]$ respectively.

\subsubsection{Algorithm performances}
In this framework, numerical solution of \eqref{sistema2Drisolto} and \eqref{sistema1D} is  relatively fast. If we consider the FREI regime in a 2D simulation at $U_0=25[cm/s]$, a 2 thread run in a laptop with dual core processor of $2.26 [GHz]$, requires approximately $4500 [s]$ to simulate an ignition/extinction cycle when an equally spaced  mesh with $dz=1.25\cdot10^{-3}[cm]$, $dr=3.225\cdot10^{-3}[cm]$ is considered and $dt=0.6\cdot10^{-6}[cm]$.\\
If a moving mesh method with $c_f=0.05$ is considered, keeping the same spatial resolution around the flame, the computational time per FREI cycle is of the order of $220[s]$. The run time can be further reduced. Since the mesh is considerably coarser, when the flame is not present within the numerical domain, the time step of the Runge-Kutta method can be adjusted at run time, so that the simulation of a FREI cycle reduces to approximately $100[s]$ maintaining a comparable accuracy of the solution.    

\section{Calibration of the theoretical models}
Let consider a pipe with a fixed diameter $d=0.2 [cm]$. We want to find
the best values of $D$ and $A^*$, within a theoretically consistent range \cite{westbrook1981simplified,cussler1997diffusion}, that 
give a response of the dynamic system which agrees as much as possible with experimental results of Tsuboi et al. \cite{Tsuboi09} in the relevant range $U_0 \in [1,100] cm/s$.
In particular, experiments  show evidence of at least 3 different regimes with unstable transitions: i) stable, ii) unstable FREI, iii) stable weak.
More in details, see Fig. \ref{fig:tutti2D}, experimental data show that for a relatively high flow rate the mixture
 is auto-ignited 
downstream (hot side of the tube) and the energy released by combustion
 is sufficient to sustain a strong stable flame that stabilize in a certain
 position of the tube. Such behavior has been observed, for a stoichiometric
 methane/air mixture, in a range of $U_0$ approximately greater than $40 cm/s$.
For values of $U_0$ between $\approx 40 cm/s$ and $ \approx 5 cm/s$, 
the FREI regime is observed. Here, the heat released during combustion does
 not allow stabilization of the flame and the reaction
 is repetitively ignited downstream and thermally quenched upstream. 
For very low flow rates ($U_0\simeq 5 cm/s $) the incoming fuel is not
 sufficient to guarantee a complete autoignition. The fuel oxidation take
 place in a stable position in the heated side of the tube and a weak peak
 of heat and mixture temperature is observed. A diffuse weak flame is thus observed.

\subsection{Calibration of 2D Model }
In principle, both $D$ and $A^{*}$ could be roughly estimated. The mass diffusivity of methane in air, for example, may be computed from the Chapman-Enskog relation \cite{cussler1997diffusion} which gives a value of $D^{0}_{CH_{4}}\approx 0.2 [cm^{2}/s]$ at $T=298 [K]$.   Nevertheless,  we are trying to represent the diffusion of a multicomponent mixture by one constant diffusion coefficient only and with the assumption of Lewis number equal to one.  Moreover, since mass diffusivity increases with temperature, a value of $D$ higher than $D^{0}_{CH_{4}}$ is expected. Even the value of the rate $A^{*}$ can be estimated from literature. Westbrook et al.~\cite{westbrook1981simplified}, for a global kinetic mechanism but with multicomponent diffusion, propose a value  $A=1.3\cdot10^{8}$ (defined in Eq.~\ref{rateWest})  which provides an estimation of $A^{*}=3.2\cdot10^{8}$. Starting from these two references, the model has been calibrated with the following requirements:
\begin{itemize}
\item First of all, to have a good representation of the velocity $U_{0}$ at which the transition from stable to FREI regime occurs;
\item second, to qualitatively match, above and near transition, the position at which the flame stabilizes;
\item third, the excursion of the flame from ignition point to the extinction point should be comparable between numerical  and experimental results.
\end{itemize}
On this basis, optimal parameters turned out to be
 $A^*=1.455\cdot10^9[K/s]$, $D=0.6667[cm^{2}/s]$,
which are physically sound for a methane/air stoichiometric mixture.  

\begin{figure}[!ht]
\centering{
\includegraphics[scale=0.4]{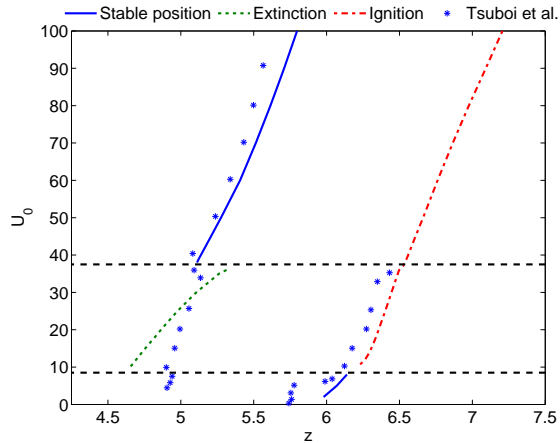} 
}
\caption{Comparison of numerical flame position (2D model) with experimental results of Tsuboi et al.\cite{Tsuboi09}. 
($D=0.6667[cm^2/s]$, $A^*=1.455 \times 10^9[K/s]$). The two horizontal dashed lines mark the unstable region.
 (Continuos line) Position at which stabilization of flame occurs. (Dash-Dotted line) Position at which ignition occurs ($\bar{Q}>1\times 10^5 [K/s]$).(Dashed line) Position at which extinction occurs ($\bar{Q}<1\times 10^5 [K/s]$).  See Eq.~\eqref{eq:medie} for the definition of $\bar{Q}$. }
\label{fig:tutti2D}
\end{figure}

Results of 2D simulations are shown in figure \ref{fig:tutti2D}, where  the 
ignition and the stabilization (or extinction) position of the flame for
 the complete range of $U_0$ are shown, together with the experimental results.
  From a numerical point of view, a flame is considered to be ignited if the mean heat release, defined in Eq.~\eqref{eq:medie}, is $\bar{Q}>1\cdot10^5 [K/s]$. On the other hand, a flame is said to be extinct when $\bar{Q}<1\cdot10^5 [K/s]$.

From the figure,  it is possible to see that all regimes are qualitatively well reproduced. Moreover, a good quantitative agreement is found in the whole range of $U_{0}$. 
Yet, some discrepancy is present for very low velocities, when weak flame is triggered and diffusion plays a central role.

\begin{figure}[!h]
\centering{
\includegraphics[scale=0.4]{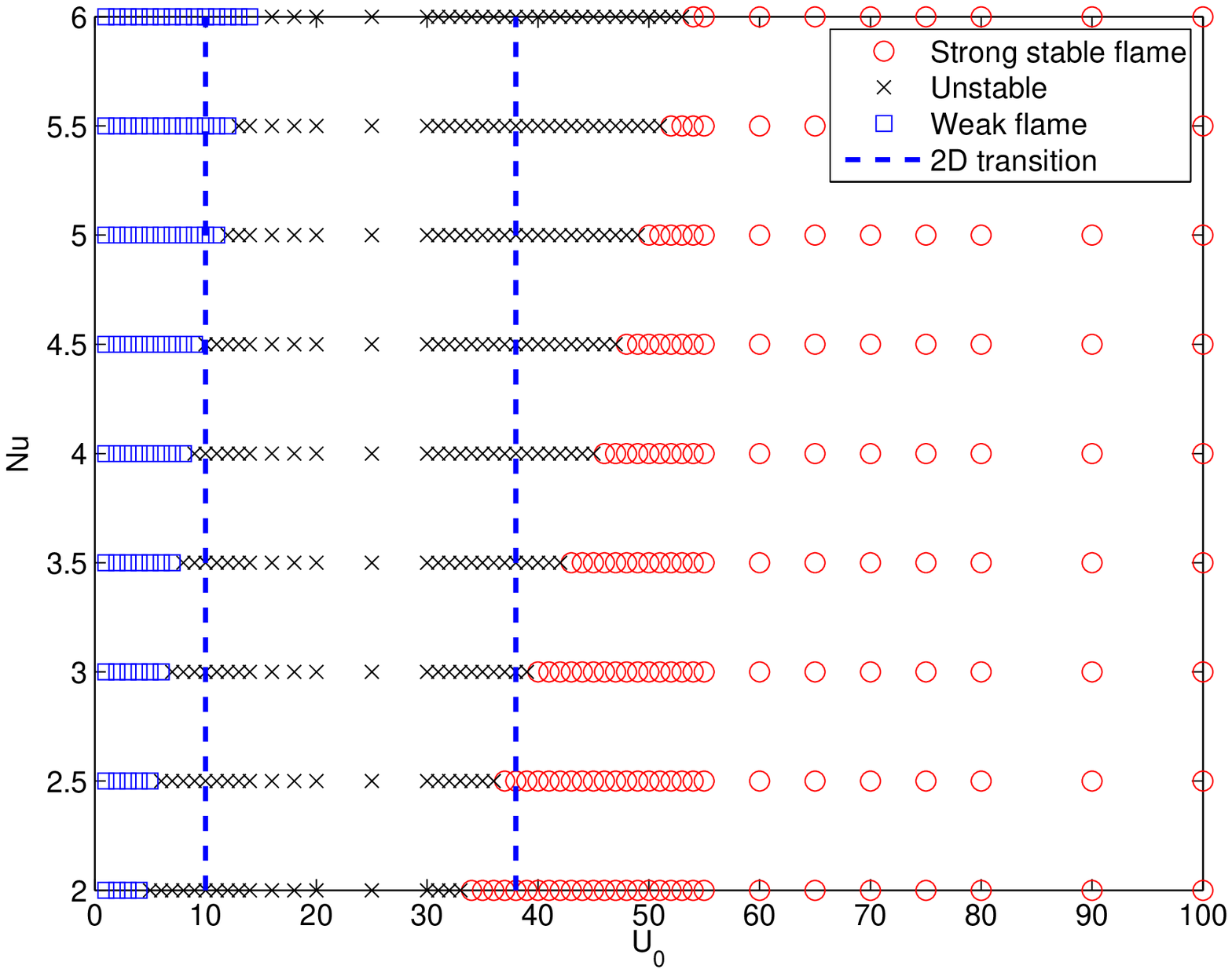} 
\includegraphics[scale=0.5]{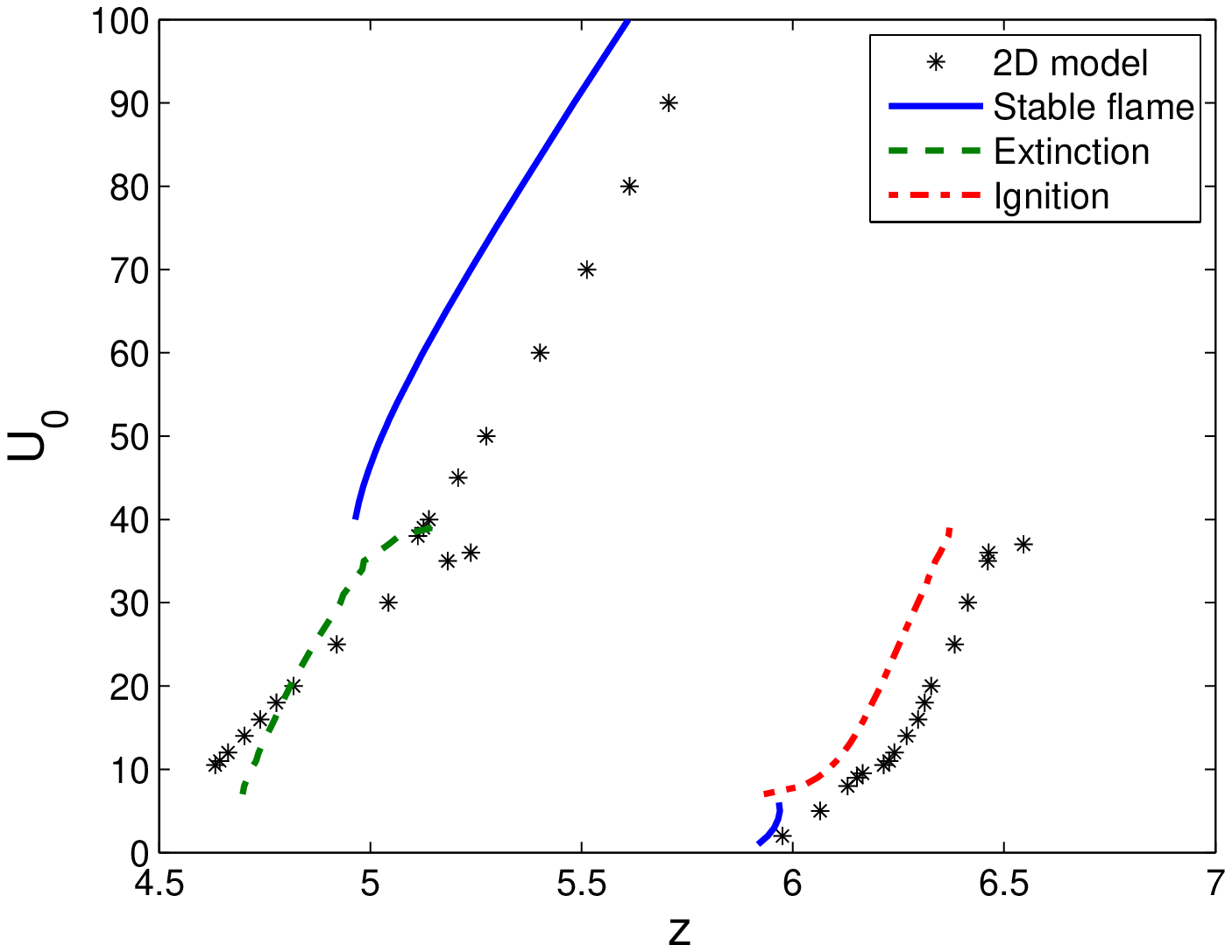} 
}
\caption{(a): Stability analysis of the microcombustor in 1-d model as a function of Nusselt and velocity. The dashed lines indicate the transition between stable and unstable regimes found in the 2-D model. The three regimes typical of this configuration are found for all the Nusselt considered here. However, the transition occurs at different velocities. We consider that the dynamics is optimally captured for $Nu=3$.
(b): 1-D model results at Nu=3. The numerical results are compared with the 2-D calculations. Small differences are present in the weak regime. The ignition and extinction positions are different in 1-D and 2-D models, remaining in qualitative agreement. Bigger differences are present for the stable branch.}
\label{fig:stabilitaNu1D}
\end{figure}   
\subsection{Calibration of 1D Model }
A 1-D model can be justified if a clear separation of scales is present\cite{sanchez,CFLV,bender,barenblatt}. 
In particular, this assumption may seem justified when the ratio between the radial and the axial length tends to zero. 
As done for 2D simulations, we solve equations (\ref{sistema1D}) for different 
values of $U_0$ in the range of $[1,100] cm/s$.
In order to calibrate the 1-D model,
 the 2-D model is taken as reference.
 Hence, the same parameters are used:
 resolution of the grid along $z$, time step,
 diameter of the tube ($d=0.2 [cm]$), wall temperature  and initial conditions.
 In this framework, the only free-parameter left is the Nusselt number and, in order to keep the model as simple as possible, we consider it  as a constant within the range of $U_{0}$ tested. \\
The Nusselt number can be roughly estimated~\cite{bookNusselt} considering that for a fully developed laminar (non-reactive) flow,
 with fixed wall temperature, the Nusselt is a constant: $Nu=3.66$.
Moreover, if the heat flux at the wall is constant, then $Nu=4.36$.
In their numerical study, Nakamura et al.~\cite{nakamura12}  showed that a FREI regime can be qualitatively reproduced with an intermediate value of Nusselt 
($Nu=4$).
In order to fix the optimal value of this parameter we 
have performed an analysis of the stability and the dynamic response of the system in the range of $Nu\in[2-6] $. Results are shown in figure \ref{fig:stabilitaNu1D}a.
One can observe that, for a small pipe of diameter $d=0.2 cm$ in the range of $U_0 \in 1 \div 100 cm/s$ , 
a 1D model is capable to reproduce qualitatively all the regimes that have been observed within the 2D simulation, 
in the whole range of Nusselt tested. 
Nevertheless, best agreement, in terms of range
 of stability and dynamic response of the system, is found for $Nu
 \approx 3$, see figure \ref{fig:stabilitaNu1D}a. 
 
 The complete dynamical analysis  is presented for this Nusselt number in Fig \ref{fig:stabilitaNu1D}b, where  the 1D and 2D results are compared in
 the whole range of $U_0$.  While the range of stability is close
 to that obtained in 2D, the results are quantitatively different, 
 even though the diameter of the pipe is 
small with respect to the axial characteristic length (or the characteristic length of the FREI oscillations).
That is particularly clear in the stable regime at high flow rates, where significant differences
 in the position at which flame stabilizes, are found.
In the FREI regime, most of the differences are in terms of location at which ignition and extinction  occur.

\section{Physical analysis in 2D} 
In the previous section we have shown that both, 1D and 2D constant density models with a global reaction mechanism, are capable to reproduce the main features of the complex dynamics involved in micro-combustors.
Since the model used is simple, it permits to easily 
point out which mechanisms are responsible for the different dynamical features.
In this section, we discuss in details the combustion dynamics of the 2D model  for a micro-channel  of diameter $d=0.2 [cm]$ at different inflow regimes.
\subsection{Stable flame regime}
\begin{figure}[h]
\centering{
\includegraphics[scale=0.5]{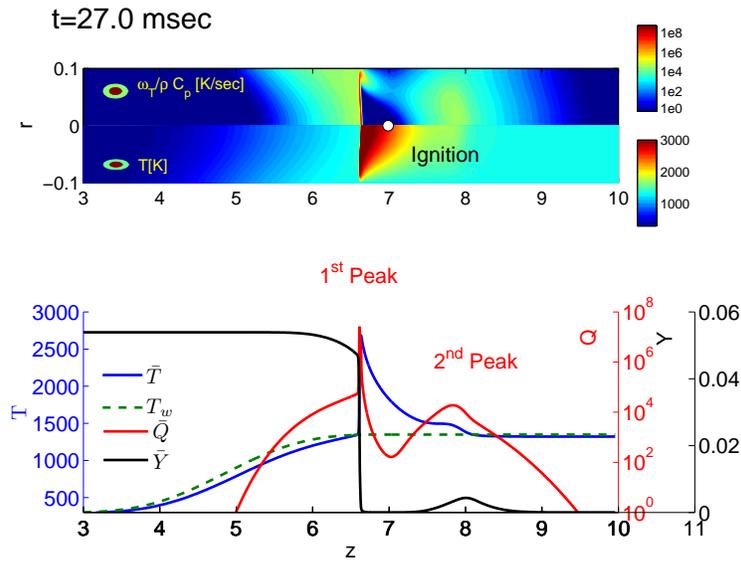}}
\caption{(Top panel) Snapshot of 2D reaction rate ($\frac{\dot\omega_{T}}{\rho C_p}$) and temperature ($T$) field at $t=0.027[s]$. The white dot shows the position ($z(\bar{Q})$) at which ignition of the flame has occurred.
 (Bottom panel) Instantaneous averaged fields. 
At this instant two peaks of $\bar{Q}$ are present. First peak is due to the stronger flame
while the second weaker peak is due to the splitting of the flame after ignition.  }
\label{fig:SnapStrong}
\end{figure} 
We describe now the dynamics of ignition in the stable regime and we consider  a value of $U_0=80 cm/s$. 
For the sake of convenience, we impose an initial condition in which temperature is constant and equal 
to $T(z,r,0)=300 K$ over the entire domain. The mass fraction of methane is set
 $Y_{CH_4}(r,z,0)=0.055$ if $z<5$, and exponentially decaying to zero if $z>5$. Such initial condition is chosen to reduce the transient before autoignition. In this way, we avoid simulating the advection-diffusion of fuel from the tube 
entrance to the hot region. 
While other initial conditions may be considered,
we have however verified that the relevant observables, as the position 
at which ignition occurs and the position where the flame stabilizes, are
 quite robust and not influenced by initial conditions.   
To analyze the dynamics of the system it is convenient
to refer to radial averaged quantities and we define the mean burning
 rate, temperature and mass fraction as follows:
\begin{align}
\bar{Q}(z,t)&=\frac{2\pi}{A}\int_{0}^R S_T r dr\nonumber \\ 
\bar{T}(z,t)&=\frac{2\pi}{A}\int_{0}^R T  r dr\nonumber \\ 
\bar{Y}(z,t)&=\frac{2\pi}{A}\int_{0}^R Y_{CH_4} r dr  
\label{eq:medie}
\end{align}
 All these variables are sampled during simulations as well
 as the position ($z(\bar{Q})$) and the peaks
 of the mean burning rate $\bar{Q}$. 
 Indeed, $\bar{Q}$ may show several peaks in the domain of different magnitude.
 For the sake of clarity, we number them in order of decreasing magnitude.
 Moreover, from the time history of the position, we can define
 the relative flame speed as $V(\bar{Q})=\frac{d z(\bar{Q})}{dt} -U_0$.\\
A snapshot of the 2D solution and the computed averaged quantities are shown in Fig.~\ref{fig:SnapStrong}. 
\begin{figure}[h]
\centering{
\includegraphics[scale=0.525]{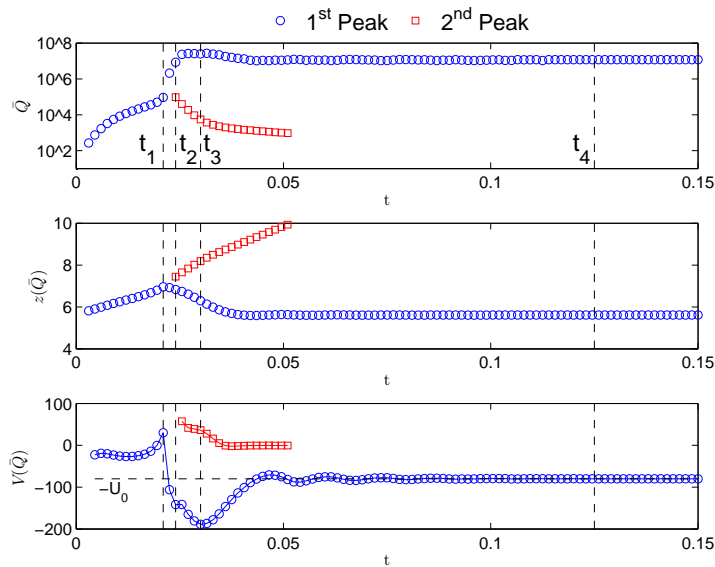}
}
\caption{Strong stable flame regime for 2D model and $U_0=80[cm/s]$. 
(Top panel) Time history of the magnitude of the peaks 
of the averaged burning rate ($\bar{Q}$). Instants $t1$, $t2$ and $t3$  are 
respectively the time of ignition, splitting and maximum speed of propagation of the primary flame. (Central panel) Position of the peaks.
 (Bottom panel) Relative speed of propagation computed as $V(\bar{Q})=\frac{d z(\bar{Q})}{dt} -U_0$.  }
\label{fig:strongQXV}
\end{figure}  
 To describe the combustion dynamics, in this regime of flow rate, we identify
 four important instants $t_1\div t_4$ (figure \ref{fig:strongQXV}). 

For $t<t_1$ a single and broad peak of $\bar{Q}$ is moving downstream
 and fuel starts being slowly consumed. Around $t_1$ there is a rapid increasing of the burning rate and the temperature of
 the mixture rises up. We define $t_1$ as the time at which ignition occurs and a strong flame is
 generated.  From a numerical point of view this event is identified when the mean reaction rate is greater than a given threshold, which in our computations has been chosen as  $\bar{Q}= 10^5 K/s$. 
 After ignition, the speed of the flame is greater than $U_0$, thus 
the peak of burning rate that was moving downstream reverses and 
starts moving upstream, see figure \ref{fig:strongQXV}b.
Around $t_2$ a second peak appears and the flame splits into two. The stronger
 flame accelerates upstream reaching a relative maximum flame speed ($t_3$) of the order of $200 cm/s$. The remaining part of fuel, 
which is not yet burned, is consumed by a second weaker flame that moves downstream,
see figure \ref{fig:strongQXV}c.
   
 At $t>t_4$ steady state is reached, the weak flame blows off the domain
 and a single slightly V-shaped flame stabilizes.
 
\subsection{Transition from stable flame to FREI }
\begin{figure}[h]
\centering{
\includegraphics[scale=0.425]{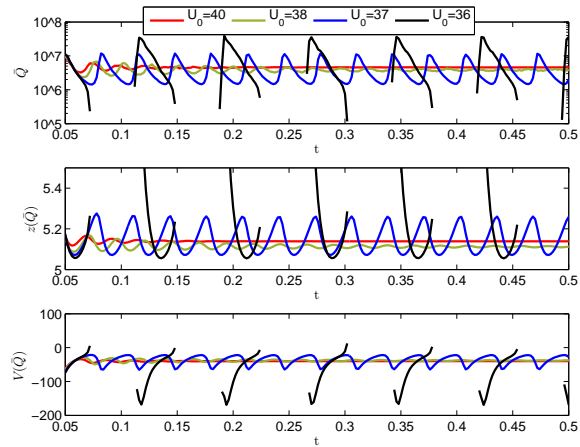}
}
\caption{(Top panel) Time history of the magnitude of the first peak of
 the averaged reaction rate, for different inflow conditions($U_0$).
For the sake of clarity, only the first peak of $\bar{Q}$ is followed in 
time here, and only if a strong flame is ignited ($\bar{Q}>10^5 [K/s]$). (Central panel) Position of the peaks and relative speed of propagation.
 ( Bottom panel).  }
\label{fig:transUP}
\end{figure}  
By reducing the inflow velocity, a transitional regime appears. In Fig.~\ref{fig:transUP} we show the dynamic response of the system
to different values of $U_0$.
While for $U_0$ sufficiently high, the stabilization of the flame
 is a very fast process that takes place in some tenths of a second, for $U_0$ approaching the
 limit of $\approx 38 cm/s$ the stronger flame shows an oscillatory behavior.
If $U_0$ is high enough ($U_0>37 cm/s$), oscillations are progressively dumped and eventually
the flame stabilizes. For $U_0$ slightly smaller than $37 cm/s$, 
these oscillation are not dumped and a pulsating flame 
is observed:
 the flame remains ignited ($\bar{Q}>10^5 K/s$) but it shows a pulsating intensity.
 The characteristic length of the oscillations is of the order of some millimeters. The frequency of pulsation is computed to be in the range $25-30 [Hz]$.
  Pulsating flames, between stable and FREI regime, have been already documented in literature for different reactive mixtures\cite{Oshibe,maruta05}. 
  Nevertheless, in the present set-up, the range of flow rate at which this phenomenon is numerically observed is very small, and therefore it seems hard to find experimental evidence of such a regime.

\subsection{Flame with repetitive ignition/extinction }
\begin{figure}[h]
\hspace{-1cm}
\centering{
\includegraphics[scale=0.425]{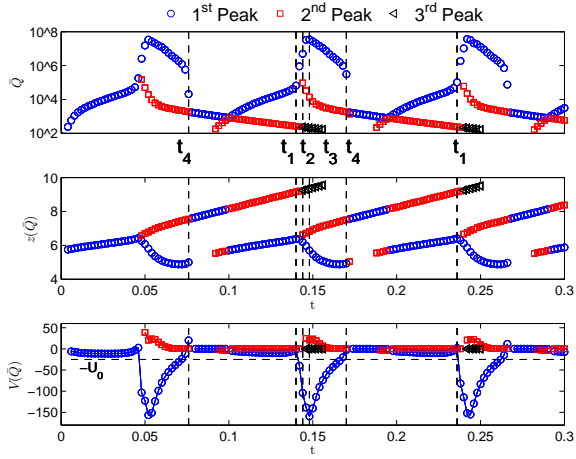}
}
\caption{Regime of flame with repetitive ignition and extinction: 2D model, $U_0=25[cm/s]$.
 (Top panel) Time history of the magnitude of the peaks 
of the averaged burning rate ($\bar{Q}$). Instants $t1$, $t2$ , $t3$ and $t4$  are 
respectively the time of ignition, splitting, maximum speed of propagation and extinction of the primary flame. (Central and bottom panel)
 Position of the peaks and relative speed of propagation.   
}
\label{fig:FREIQX}
\end{figure}
To describe the dynamics of the FREI regime, we consider $U_0=25 cm/s$ which is roughly in the middle of the velocity region 
where FREI has been observed experimentally. 
 Apart from the first cycle, which is slightly influenced by initial conditions, at this flow rate the model
 exhibits a periodic dynamics.
The time history of $\bar{Q}$ peaks is shown in figure \ref{fig:FREIQX}. 
It is possible to identify four important steps $t_1-t_4$.\\
Before $t_1$ and after $t_4$ fuel is consumed at a very low rate. 
We can identify two small peaks of the mean burning rate moving downstream. The first peak
is due to the consumption of the new fresh material coming from the tube entrance.
The second, instead, is originated from
the splitting of the flame of the previous cycle,
and it moves downstream until it reaches the end of the
 computational domain.
\\From $t_1$ to $t_2$ ignition of the strong flame ($\bar{Q}>10^5 K/s$) and flame splitting occurs.\\ 
From the splitting of the flame, after $t_2$, three distinct peaks of $\bar{Q}$ are present.
The second and the third peak of $\bar{Q}$ move downstream, 
and are the weak flames generated in the present and in the previous cycle respectively. 
On the contrary, the first peak is due to the strong flame that is moving upstream.
During the propagation, at $t_3$, for $U_0=25 cm/s$ the stronger flame reaches
 a maximum velocity of the order of $150 cm/s$ which 
 is in reasonable agreement with typical flame speeds observed for stoichiometric methane/air mixtures at atmospheric pressure. 
Nonetheless, to the best of our knowledge, experimental measurements of the flame speed for this kind of micro-channel set-up are not available in literature. \\
At $t_4$  the flame reaches a position at which the heat loss through walls is such that the 
reaction is not stable.
The flame rapidly extinguishes (the magnitude of $\bar{Q}$ decreases rapidly under the threshold) and the cycle restarts.\\
The frequency at which ignition-extinction phenomenon occurs, 
in the range of  $U_0 \approx 10-40 cm/s$, is $f\approx 5-15 [Hz]$. 

\subsection{Weak flame regime }
\begin{figure}[h]
\centering{
\includegraphics[scale=0.4]{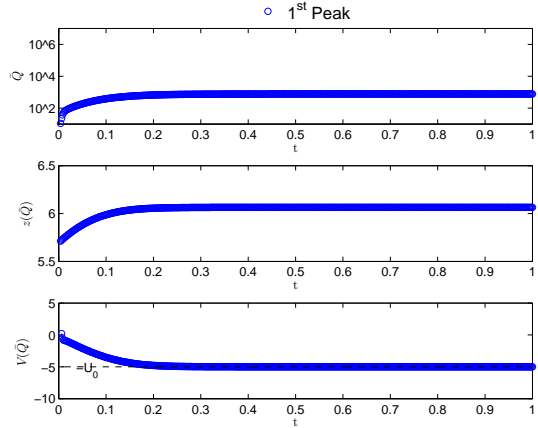}
}
\caption{Magnitude, position and relative speed of the primary peak of $\bar{Q}$ at $U_0=5[cm/s]$.}
\label{fig:weak}
\end{figure}

In the case of very low flow rate, $U_0<10 cm/s$, the model shows again stable solutions.
The time history of the peaks of the mean burning are shown in Fig.~\ref{fig:weak} for $U_0=5[cm/s]$.
In this regime the dynamics is simple.
 The incoming fresh fuel is not
 sufficient to guarantee the ignition of a strong flame.
A small and broad peak, whose magnitude is 
always under the ignition threshold of $\bar{Q}<10^5$, flows downstream until it stabilizes  at a given position. 
Fuel is partially consumed and the mean temperature of the mixture is slightly higher than the wall temperature.  

\section{Sensitivity to the diameter}
In the previous sections, the numerical models 
have been calibrated and compared against experimental results\cite{Tsuboi09}.
It has been found both qualitative and quantitative agreement within the whole range of experimental results.

In this section, we want to investigate to what extent the 2D and 1D models differ and in which way. This is important to evaluate the modeling capabilities and limits of each framework. In particular, we want to determine if there is a sensitivity to the diameter of the tube,
notably if  stability is affected by these changes.
To this aim, we have carried out simulations with varying inflow conditions at different diameters of the tube.
\begin{figure}[!ht]
\centering{
\includegraphics[scale=0.35]{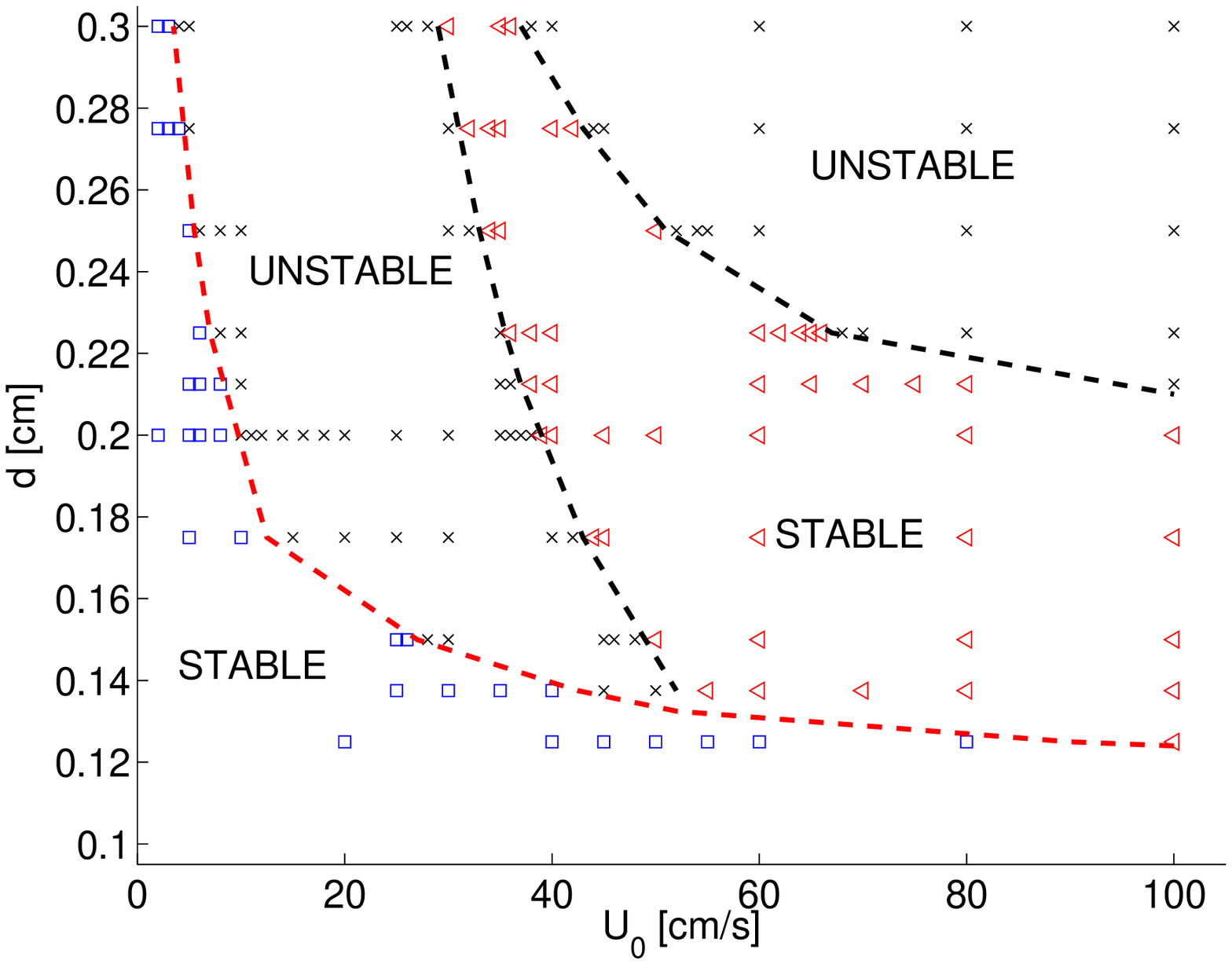} 
\includegraphics[scale=0.35]{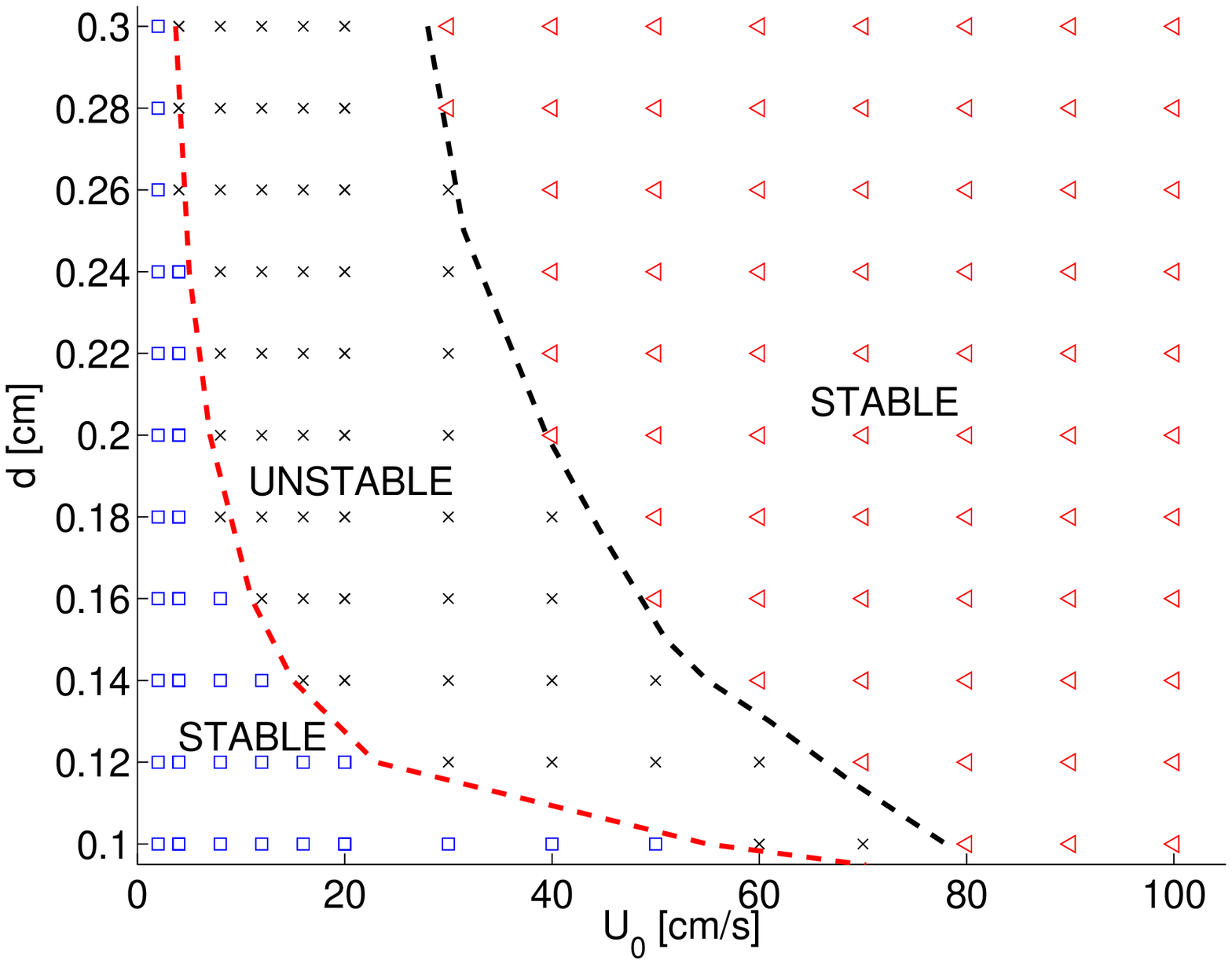} 
}
\caption{Stability analysis with respect to diameter and velocity of (a) the 2D model, (b) the 1-D model.The symbols are colored and chosen as follows: (Red square) Strong stable flame; (black X) Unstable (Blue square) stable Weak flame.}
\label{fig:stabilita12D}
\end{figure}
Results for the 2-D model are shown in Fig.~\ref{fig:stabilita12D}a, and the corresponding results for the one-dimensional model at $Nu=3$ are shown 
in Fig.\ref{fig:stabilita12D}b.
The horizontal line at $d=0.2 cm$ represents the results previously analyzed.\\
By comparing the two figures, some comments are in order:
\begin{enumerate}
\item[(i)] The three regimes observed in the previous sections (strong stable, FREI unstable, weak stable) are present in both simulations;
\item[(ii)] The transition line between weak stable regime and FREI 
is not much sensitive to the dimensionality of the model for diameters of the tube between $0.175-0.3$ [cm]. Simple physical reasoning suggests  that:  for a fixed and low $U_{0}$ the magnitude of the velocity gradient along $r$ is smaller when  $R$ is increased, and  diffusion plays an important role when $U_{0}$ is small. Therefore, a 1D approximation can be sufficient to correctly mimic the mechanism of this instability.
\item[(iii)] For diameters smaller than $0.175$ [cm] the 1D model fails to reproduce correctly the transition weak-FREI.  In particular, for a  fixed diameter, the velocity at which the transition occurs is underestimated. Here, 1D  model underestimates the heat loss through the walls.  This consideration is supported by the fact that an increase of the Nusselt number (at fixed diameter) leads to a shifting of the transition toward greater values of $U_{0}$ (see Fig~\ref{fig:stabilitaNu1D}a);
\item[(iv)] Similarly, the second transition between unstable FREI and stable strong regime is well captured for higher diameters $d>0.15 [cm]$,
whereas for smaller diameters the differences are  remarkable. In general, the region of the plane $U_{0} - d$ where FREI regime is observed is largely overestimated by the 1D model.
\item[(v)] Combining a high flow rate with a bigger diameter, $d>0.21\textrm{cm}, U_{0}>40\textrm{cm/s}$, a new region of instability is experienced within the 2D framework. This region is not at all captured by the 1D model.
\end{enumerate}
\begin{figure}[h]
\centering{
\includegraphics[scale=0.425]{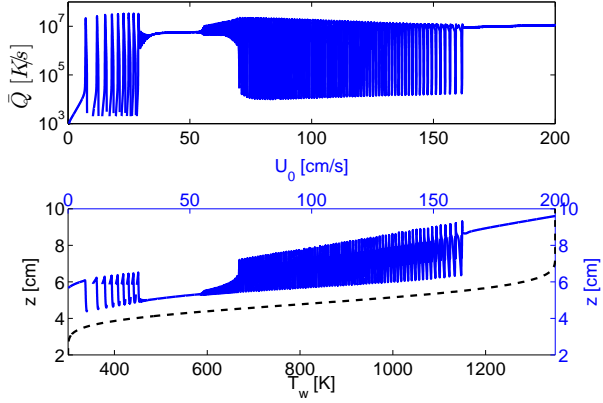} 
}
\caption{In continuos blue line we show the position of the first peak of reaction in a tube of diameter $d=0.275 [cm]$. We impose $U_{0}=0$ at $t=0$ and the flow is accelerated in time with a linear law $U_{0}=20t$. In black dashed line the temperature imposed at the solid walls. The following regimes can be observed by increasing $U_{0}$:  stable weak flame, FREI, stable strong flame, pulsating flame,FREI, stable strong flame.}
\label{fig:doubleFREI}
\end{figure}
Generally speaking, the 2-D simulations show a richer dynamic at higher diameters. In the top-right of Fig.~\ref{fig:stabilita12D}a an unstable branch appears for high-velocity flow together with a diameter large enough. A similar phenomenon has been observed by Kurdyumov et al. \cite{kurdyumov2009dynamics} in a planar geometry and 2D thermo-diffusive model.  In Fig.~\ref{fig:doubleFREI} we show the results of  a simulation for a micro-tube of diameter $d=0.275 \textrm{cm}$ in which, starting from $U_{0}=0$, the mixture is accelerated very slowly up to $U_{0}=200 \textrm{cm/s}$ (see also the supplementary material for the complete corresponding video). For low flow rates a weak flame is observed. The heat released is  small and increases with increasing $U_{0}$.  By accelerating the flow the first strong ignition occurs in correspondence with the first peak of $\bar{Q}$ in the upper panel of Fig.~\ref{fig:doubleFREI}.  The peak is followed by  a rapid decreasing of the heat release which corresponds to the thermal quenching of the flame in the cold side of the tube. In this range of $U_{0}$, the unstable FREI regime is observed. At higher $U_{0}$ ignition/extinction phenomena is suppressed and the flame remains ignited while moving downstream. If we keep accelerating the flow, the flame starts  destabilizing again and a new regime appears. The flame is ignited but it shows a pulsating heat-release combined with rapid space oscillations.  The amplitude of oscillations grows with increasing $U_{0}$ until a new FREI-like regime appears. Finally,  for $U_{0}$ sufficiently high, the flame stabilizes again and flows regularly downstream. 
Let us consider again Fig.~\ref{fig:stabilita12D}b, it is clear that the 1D model does not capture the second  unstable regime, at higher diameters.
\begin{figure}[!ht]
\centering{
\includegraphics[scale=0.425]{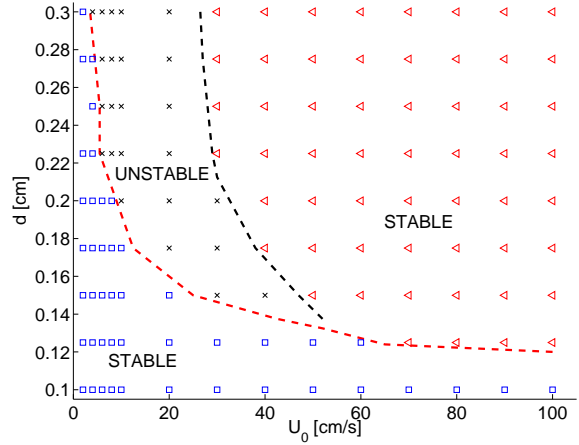} 
}
\caption{Parametric stability analysis with respect to diameter and velocity of a 2D model with flat profile velocity}
\label{fig:stabilita15D}
\end{figure}

In order to trace back this relevant difference to some physical mechanism, we have devised a thought experiment:
 a 2-D tube with a 1-D flat velocity profile.
The numerical simulations of such test-case are shown in Fig~\ref{fig:stabilita15D}.
In this case, the equations are the same as in the full 2-D, that is the heat loss through the wall is simulated without recurring to any model. Nevertheless, the velocity is imposed as in the 1-D case, \emph{i.e.} velocity does not depend on $r$ and is equal to $U_{0}$.\\
One can see that, in this configuration,  results are in the middle between those of the full 2-D and the 1-D. In particular, the two transition lines between the stable and unstable regimes are near to those obtained with the full 2-D model. This means that the ability to capture these transitions is mainly related to the amount  of energy exchanged between the mixture and the solid walls. However,  the high-flow rate unstable branch is  still not present. This evidence shows that this instability is due to the velocity gradient near-to-the-wall and cannot be captured in any case within a 1-D model.
\section{Conclusions}
We have numerically studied the dynamics of micro-combustion in tubes with diameter $d$ very small compared to its characteristic length $L$. 
More specifically, we have investigated a set-up recently studied experimentally\cite{Tsuboi09}. 
In order to disentangle different mechanisms at play, 
we have used a 2D axially symmetric and a 1D thermo-diffusive model which can be derived, under suitable hypothesis, from a more detailed mathematical model. \\
Our main findings are:
\begin{itemize}
\item The important features of the combustion instabilities in microtubes
can be well described with low-order models with only one-step
chemical reaction. The complete chemistry does not appear to be crucial.
Indeed, we have shown that by considering a set of parameters which are physically sound, the numerical results are qualitatively and quantitatively in agreement with experiments within the whole range of flow-rates experimentally explored.
\item At variance with recent computational analysis\cite{nakamura12}, 
our results show that the mechanism of the FREI instability is essentially
hydrodynamical, and not related to chemistry balance.
\item In the weak-flame regime,  present approximations are too crude. Diffusion, detailed chemistry together with coupling between chemistry and hydrodynamics  are more important.
\item  We have finally shown that the dimensionality of the model can play an important role and that a 1D model should be used very carefully.  It can give some qualitative informations about the range of stability at very low computational cost, but the validity of the results are limited.
More specifically, at lower flow-rate, the 1D model overestimates the FREI branch. Instead, at higher flow-rate, it is not capable to capture 
a second unstable branch found in 2D simulations.
\end{itemize}

The application of the present low-order model to a different experimental set-up is under investigation in order to assess
present results and the model limitations.
Finally, work in progress considers linear and non-linear stability analysis of the 2D and 1-D model to shed some light on the details of 
mechanism of the instability and to explain the richer dynamics observed at higher diameters.



\begin{thebibliography}{1}

\bibitem{maruta11}
K.~Maruta,
{\em Proc. Combust. Inst. \/} 33 (2011) 125--150.

\bibitem{fernandezpello02}
A.C.~Fernandez-Pello,
{\em Proc. Combust. Inst. \/} 29 (2002) 883--899.
  
\bibitem{pizza09}
G.~Pizza, J.~Mantzaras, C.E.~Frouzakis, A.G.~Tomboulides, K.~Boulouchos,
{\em Proc. Combust. Inst. \/} 32 (2009) 3051--3058.

\bibitem{lloyd74}
S.A.~Lloyd, F.J.~Weinberg,
{\em Nature \/} 251 (1974) 47--49.

\bibitem{chen11}
C.H.~Chen, P.D.~Ronney,
{\em Proc. Combust. Inst. \/}  33 (2011) 3285--3291.

\bibitem{maruta05}
K.~Maruta, T.~Kataoka, N.l.~Kim, S.~Minaev, R.~Fursenko,
{\em Proc. Combust. Inst. \/}  30 (2005) 2429--2436.

\bibitem{pizza08a}
G.~Pizza, C.E.~Frouzakis, J.~Mantzaras, A.G.~Tomboulides, K.~Boulouchos,
{\em Comb. Flame} 155 (2008) 2--20.

\bibitem{kurdyumov02}
V.N.~Kurdyumov, E.~Fern{\'a}ndez-Tarrazo,
{\em Comb. Flame} 128 (2002) 382--394.

\bibitem{kurdyumov07}
V.N.~Kurdyumov, E.~Fern{\'a}ndez-Tarrazo, J.-M.~Truffaut, J.~Quinard, A.~Wangher, G.~Searby,
{\em Proc. Combust. Inst. \/} 31 (2007) 1275--1282.

\bibitem{tsai08}
C.-H.~Tsai,
 {\em Combust. Sci. Technol.} 180 (2008) 533--545.

\bibitem{dogwiler98}
U.~Dogwiler, J.~Mantzaras, P.~Benz, B.~Kaeppeli, R.~Bombach, A.~Arnold,
{\em Proc. Combust. Inst. \/} 27 (1998) 2275--2282.

\bibitem{dunn86}
D.~Dunn-Rankin, P.K.~Barr, R.F.~Sawyer,
{\em Proc. Combust. Inst. \/} 21 (1988) 1291--1301.

\bibitem{bychkov07}
V.~Bychkov, V.~Akkerman, G.~Fru, A.~Petchenko, L.-E.~Eriksson,
{\em Comb. Flame} 150 (2007) 263--276.

\bibitem{richecoeur05}
F.~Richecoeur, D.C.~Kyritsis,
{\em Proc. Combust. Inst. \/} 30 (2005) 2419--2427.

\bibitem{jackson07}
T.L.~Jackson, J.~Buckmaster, Z.~Lu, D.C.~Kyritsis, L.~Massa,
{\em Proc. Combust. Inst. \/}  31 (2007) 955--962.

\bibitem{kurdyumov08}
V.N.~Kurdyumov, J.-M.~Truffaut, J.~Quinard, A.~Wangher, G.~Searby,
{\em Combust. Sci. Technol.} 180 (2008) 731--742.

\bibitem{kwon96}
M.J.~Kwon, B.J.~Lee, S.H.~Chung,
{\em Comb. Flame} 105 (1996) 180--188.

\bibitem{xu07}
B.~Xu, Y.~Ju,
{\em Proc. Combust. Inst. \/} 31 (2007) 3285--3292.

\bibitem{fan09}
Y.~Fan, Y.~Suzuki, N.~Kasagi,
{\em Proc. Combust. Inst. \/} 32 (2009) 3083--3090.

\bibitem{pizza10}
G.~Pizza, C.~E. Frouzakis, J.~Mantzaras, A.G.~Tomboulides, K.~Boulouchos,
{\em J. Fluid Mechanics} 658 (2010) 463--491.

\bibitem{yamamoto11}
A.~Yamamoto, H.~Oshibe, H.~Nakamura, T.~Tezuka, S.~Hasegawa, K.~Maruta,
{\em Proc. Combust. Inst. \/} 33 (2011) 3259--3266.

\bibitem{nakamura12}
H.~Nakamura, A.~Fan, S.~Minaev, E.~Sereshchenko, R.~Fursenko, Y.~Tsuboi, K.~Maruta,
{\em Comb. Flame} 159 (2012) 1631--1643.

\bibitem{Tsuboi09}
Y.~Tsuboi, T.~Yokomori, K.~Maruta,
{\em Proc. Combust. Inst. \/} 32 (2009) 3075--3081.

\bibitem{poinsot}
T.~Poinsot, D.Veynante,
\newblock {\em Theoretical and numerical combustion, 2$^{\mathrm{nd}}$ edition \/},
\newblock Edwards ed., 2005.

\bibitem{bai13}
B.~Bai, Z.~Chen, H.~Zhang, S.~Chen,
\newblock {\em Comb. Flame} 157 (2013) 1572--1580.

\bibitem{Peters}
N. Peters,
\newblock {\em Turbulent combustion \/},
\newblock Cambridge University Press, 2000.

\bibitem{westbrook1981simplified}
C.K.~Westbrook, F.L.~Dryer,
{\em Combust. Sci. Technol.} 27 (1981) 31--43.

\bibitem{Fornberg98}
B.~Fornberg,
{\em SIAM Rev \/} 40 (1998) 685--691.

\bibitem{fornberg1988generation}
B.~Fornberg,
{\em Mathematics of Computation} 51 (1988) 699--706.

\bibitem{cussler1997diffusion}
E.L.~Cussler,
\newblock {\em Diffusion: mass transfer in fluid systems \/},
\newblock Cambridge university press, 1997.

\bibitem{Oshibe}
H.~Oshibe, H.~Nakamura, T.~Tezuka, S.~Hasegawa, K.~Maruta,
{\em Comb. Flame} 157 (2010) 1572--1580.

\bibitem{sanchez}
E.~Sanchez-Palencia,
{\em Non-homogeneous media and vibration theory},
\newblock Lecture notes in physics 127, Springer (1980).

\bibitem{CFLV}
P.~Castiglione, M.~Falcioni, A.~Lesne, A.~Vulpiani,
\newblock {\em Chaos and coarse graining in statistical mechanics \/},
\newblock Cambridge University Press, (2008).  

\bibitem{bender}
C.M.~Bender, S.A.~Orszag,
\newblock {\em Advanced mathematical methods for scientists and engineers I: Asymptotic methods and perturbation theory \/},
\newblock Springer, (1999).  

\bibitem{barenblatt}
G.I.~Barenblatt,
\newblock  {\em Scaling, self-similarity, and intermediate asymptotics: dimensional analysis and intermediate asymptotics \/}, 
\newblock Cambridge University Press, (1999).  

\bibitem{kurdyumov2009dynamics}
V.N.~Kurdyumov, G.~Pizza, C.E.~Frouzakis, J.~Mantzaras,
{\em Comb. Flame} 156 (2009) 2190--2200.



\bibitem{flamethickness}
Heravi, H. M., Azarinfar, A., Kwon, S. I., Bowen, P. J., Syred, N.
{\em Third European Combustion Meeting ECM} (2007) 1--6


\bibitem{bookNusselt}
Incropera, F. P., Lavine, A. S., DeWitt, D. P.
{\em Fundamentals of heat and mass transfer.} (2011) 


\end{thebibliography}
\end{document}